\documentclass[11pt]{article}
\usepackage{fleqn,cospar}

\usepackage{url}

\usepackage{graphicx}
\usepackage[figuresright]{rotating}

\newcommand{\xmm}{{\it XMM-Newton}}
\newcommand{\sas}{{\it XMMSAS}}
\newcommand{\epic}{{\it EPIC}}
\newcommand{\mos}{{\it MOS}}
\newcommand{\pn}{{\it pn}}
\newcommand{\bm}{$\beta$-model}
\newcommand{\tab}[1]{Table~\ref{#1}}
\newcommand{\fig}[1]{Figure~\ref{#1}}
\newcommand{\etal}{et~al.} 

\newcommand{\rs}{$R_{1\sigma}$}

\title{Measuring the matter distribution
       within $z=0.2$ cluster lenses with \xmm}

\author{P.B. Marty
        \address{I.A.S., Universit\'e Paris-Sud, B\^at. 121, 
		 F-91405 Orsay cedex, France}$^{,2}$,
	S. Bardeau
	\address{L.A.T., Observatoire Midi-Pyr\'en\'ees,
		 14 av. Edouard Belin, F-31400 Toulouse, France},
	O. Czoske$^{2}$,
	H. Ebeling
        \address{Institute for Astronomy, University of Hawaii, 2680
		 Woodlawn Drive, Honolulu, HI\,96822, USA},
	J.P. Kneib$^{2,}$
	\address{Caltech-Astronomy, MC 105-24, Pasadena, CA 91125, USA},
	R. Sadat$^{2}$,
        and
	I. Smail
        \address{Institute for Computational Cosmology, University of Durham,
		 South Road, Durham DH1 3LE, UK}}

\begin{document}

\maketitle

\begin{abstract}
We present an analysis of $7$ clusters observed by \xmm\ as part of our survey
of $17$ most X-ray luminous clusters of galaxies at $z \sim 0.2$ selected for
a comprehensive and unbiased study of the mass distribution in massive
clusters. Using the public software {\it FTOOLS} and \sas\ we have set up an
automated pipeline to reduce the \epic\ \mos\ \& \pn\ spectro-imaging data,
optimized for extended sources analysis. We also developped a code to perform
intensive spectral and imaging analysis particularly focussing on proper
background estimate and removal. \xmm\ deep spectro-imaging of these clusters
allowed us to fit a standard \bm\ to their gas emission profiles as well as
a standard MEKAL emission model to their extracted spectra, and test their
inferred characteristics against already calibrated relations.
\end{abstract}

\section*{INTRODUCTION}

Being the most massive gravitationally bound objects in the Universe,
clusters of galaxies are prime targets for studies of structure
formation and evolution.

Of particular interest are the most X-ray luminous clusters.
Indeed, it has been convincingly
demonstrated that high X-ray luminosity is a reliable indicator of
cluster mass ({\it e.g.} \cite{hilx:Allen98}).
Therefore, X-ray selected clusters samples
({\it e.g.} \cite{hilx:Ebeling96}) are the best choice for
undertaking representative studies of the mass distribution in massive,
X-ray luminous clusters. In particular, as they are easily detected
out to very large redshift, we can hope to probe cluster evolution back from
early epochs. However, X-ray luminosity is also sensitive to the
presence of cooling flows, mergers and non-thermal effects, thus the 
Luminosity-Mass and Temperature-Mass relation needs to be clearly understood
before any strong ascertion is made on cluster evolution (\cite{hilx:Smith02}).
Proper mass estimate is likely to arise only by combining multiwavelength 
observations of clusters of galaxies: namely strong and weak lensing 
observations (\cite{hilx:Smith01}), 
velocity distribution and dynamics of the cluster galaxies (in a similar
way as in \cite{hilx:Czoske01}) and through accurate, spatially
resolved X-ray temperature measurements as we are concentrating on here.

Following this premise, we are conducting a survey (\tab{hilx:amasamp})
of the most X-ray luminous clusters
($L_{X_{[0.1-2.4~keV]}} \ge 8~10^{44}~erg \cdot s^{-1}$)
in a narrow redshift slice at $z\sim 0.2$,
selected from the XBACS (X-ray Brightest Abell-type Clusters Sample;
\cite{hilx:Ebeling96}). As XBACS is restricted to Abell clusters
(\cite{hilx:ACO89}), it is X-ray flux limited but not truly X-ray 
selected. However, a comparison with the X-ray selected {\it ROSAT}
BCS (Brightest Cluster Sample, \cite{hilx:Ebeling00}) shows 
that $\sim 75\%$ of the BCS clusters in the redshift and X-ray 
luminosity range of our sample are in fact Abell clusters. Hence, 
our XBACS subsample is, in all practical aspects, indistinguishable 
from an X-ray selected sample.

\begin{table*}[t]
  \vspace{-8mm}
  \begin{minipage}{170mm}
    \begin{footnotesize}
    \caption{\label{hilx:amasamp} Our cluster list sorted by right
         ascension. Redshift, Hydrogen column density, flux and luminosity
         (columns 3 to 6 resp.) are quoted from the XBACS updated database.
         Out of these 17 clusters, 8 have HST/WFPC2 observations performed
         (as part of cycle 8 [ID:8249, PI:Kneib]); 3 only have a shallow HST
         observation through a snapshot survey (8301 \& 8719 programmes,
         PI:Edge); 2 have not yet been observed with HST; the remaining 4
         having archival data. Moreover, 11 have been observed in 1999/2000
         with the wide field CFH12K camera (\cite{hilx:Czoske03}).
         The last column shows the effective
         exposure duration (after flares removal) for clusters already
         observed by \xmm\ (resp. \mos\ and \pn\ instruments), or quote the
         priority level within our proposal for those not yet observed.}
    \begin{tabular}{lccccccccc}
\hline
ID&$\alpha$&$\delta$&$z$&$N_H$&$f_{xbacs}$&$L_{xbacs}$&HST&CFH&\xmm          \\
&$J2000$&$J2000$&&$10^{20}~cm^{-2}$&$10^{-12}~cgs$&$10^{44}~erg/s$&cycle&semester&$ks$ (\mos/\pn)\\
\hline
A68  &$00~37~06$&$+09~09~20$& 0.2546 & 4.77 & 5.55 & 14.90 & 8249&99-II&22/10\\
A115 &$00~55~55$&$+26~22~14$& 0.1971 & 5.39 & 8.99 & 14.57 & -   &  -  &C    \\
A209 &$01~31~53$&$-13~36~47$& 0.2060 & 1.56 & 7.55 & 13.38 & 8249&99-II&17/11\\
A267 &$01~52~41$&$+01~00~43$& 0.2300 & 2.84 & 6.21 & 13.67 & 8249&99-II&17/12\\
A383 &$02~48~03$&$-03~31~04$& 0.1871 & 4.06 & 5.50 &  8.09 & 8249&99-II&25/07\\
A773 &$09~17~52$&$+51~43~38$& 0.2170 & 1.34 & 6.36 & 12.50 & 8249&  -  &15/15\\
A963 &$10~17~03$&$+39~02~56$& 0.2060 & 1.40 & 5.78 & 10.27 & 8249&99-I &24/16\\
A1423&$11~57~17$&$+33~36~39$& 0.2130 & 1.66 & 5.22 &  9.92 & 8719&  -  &C    \\
A1682&$13~06~49$&$+46~33~35$& 0.2260 & 1.36 & 5.44 & 11.59 & 8719&  -  &C    \\
A1689&$13~11~29$&$-01~20~31$& 0.1840 & 1.80 &14.60 & 20.64 & 6004&00-I &AO1 J.Hugues\\
A1763&$13~35~19$&$+40~59~56$& 0.2279 & 0.84 & 6.54 & 14.13 & 8249&99-II&observed\\
A1835&$14~01~02$&$+02~52~03$& 0.2528 & 2.24 &14.58 & 38.24 & 8249&99-I &PV (27/23)\\
A2111&$15~39~42$&$+34~25~03$& 0.2290 & 2.07 & 5.01 & 10.97 & -   &  -  &C    \\
A2218&$16~35~54$&$+66~13~00$& 0.1750 & 3.20 & 7.3  & 17.99 &5701, 8500&00-I&GT M.Turner\\
A2219&$16~40~20$&$+46~42~27$& 0.2281 & 1.73 & 9.23 & 19.89 & 6488&00-I &GT M.Turner\\
A2261&$17~22~27$&$+32~07~53$& 0.2240 & 3.36 & 8.72 & 18.15 & 8301&  -  &AO1 J.Hughes\\
A2390&$21~53~37$&$+17~41~45$& 0.2310 & 6.70 & 9.50 & 21.25 & 5352&00-I &GT M.Watson\\
\hline
    \end{tabular}
    \end{footnotesize}
  \end{minipage}
  \hfil\hspace{\fill}
\end{table*}

\section*{THE ANALYSIS PIPELINE}

As to the version \verb+5.2+ of the official \xmm\ data analysis software
(\sas), the extended sources could not be handled properly, regarding the
background subtraction and vignetting corrections. The
development of methods, where weights are applied to events according to
their energies and positions as well as calibration data so as to correct the
event list as if the instrument response were uniformely identical to the
on-axis response, started as soon as 2001
(\cite{hilx:Arnaud01b,hilx:Maj02,hilx:Marty02,hilx:Marty03}). We concentrated
on our side to embedding these algorithms into an environment capable of
pipeline processing a series of data from clusters of galaxies observations.
Indeed, the manual operation of \sas\ (using either external correction
routines or integrated algorithms from versions \verb+5.3+ and later) is
extremely time-consuming: several days for each dataset.

Thanks to our pipeline, we reduced the required time down to less than $10~h$
per dataset, adding a working day for eye-inspection of the full sample results
and another for compilation and final manual calculations.

This pipeline integrates the latest calibration data from the \epic\ consortium
as well as specific fitting routines written in IDL and XSPEC environments. It
is fully configurable to allow eye-inspection and partial or total re-runs
if needed. It is extensively detailed by \cite{hilx:Marty03}, but useful
information about ``blank'' and ``dark'' fields (for background estimation and
subtraction) are also respectively given by \cite{hilx:Lumb02,hilx:Marty02}.
The general processing scheme is the following:

\begin{tabular}{cl}
$\bullet$ & re-process each cluster raw data to account for latest calibration knowledge; \\
$\bullet$ & re-project ``blank'' and/or ``dark'' data onto telescope aspect solution used for each cluster; \\
$\bullet$ & extract valid events within good-time-intervals (to reject noise and proton flares); \\
$\bullet$ & detect and mask out spurious components (field sources and remaining bad pixels); \\
$\bullet$ & weight the events according to calibration data; \\
$\bullet$ & find the best cluster centroid; \\
$\bullet$ & process radial brightness profiles and fit a standard \bm\ convolved by the instrument PSF; \\
$\bullet$ & process full maps in physical flux units and adaptively smooth according to PSF description; \\
$\bullet$ & process spectra and call XSPEC for fitting (global for the mean temperature, annuli for the profile); \\
$\bullet$ & compute hardness maps and derive another temperature profile.\\
\end{tabular}

\section*{THE PIPELINE RESULTS}

\subsection*{Surface brightness profiles}
The core radius $R_c$, power index $\beta$, normalization $f_0$ and background
constant $BKG$ are coming from the standard \bm\ fitting. Note that the
surface brightness profiles have been extracted between $0.3$ and $6.3~keV$,
logarithmically rebinned to enhance their signal-to-noise ratios and that the
inner $40~arcsec$ of each cluster profile have been ignored for the fit so as
to avoid fitting any emission excess, associated for example to a cooling flow.
The background subtraction concerned only the instrumental component so that
the $BKG$ constant is a measure of the sky X-ray emission background. The
\bm s are convolved by the instrument PSF before fitting.

The \rs\ parameter is a measure of the radius inside which the
cluster signal-to-noise ratio is greater than $1$. The $F_X(data)$ and
$F_X(model)$ parameters then
are the integrated fluxes within this detection limit radius, from the data
and from the best fit \bm; in each case, these fluxes are cleaned from the
sky background component; the difference between them thus shows the
contribution of any excess emission.

\subsection*{Surface brightness maps}
Surface brightness maps that have been produced in the same energy band during
the pipeline analysis are not presented herein, because of room restrictions
and because they are not necessary for the following calculations. However,
they may be found on the internet\footnote{Only compiled on 
\url{ftp://www-station.ias.u-psud.fr/pub/epic/HiLx/2002Marty_santiago_poster.jpg}
at the moment, but individual maps will soon be available in the same ftp
directory.} if so desired.

\subsection*{Spectral analysis}
Spectra are extracted in the $0.2$ to $12.0~keV$ band and rebinned according
the Poisson statistics requirements (at least $25$ net counts per bin). They
are background subtracted from both instrumental components (estimated
using the ``closed'' data) and sky components (from a spectrum extracted in
a peripheral annulus of each dataset, between the $600$ and $800~arcsec$ radii)
and then fitted against the $TBabs \times MEKAL$ photo-absorption and
bremsstrahlung models.

The redshifts $z$ and hydrogen column densities $N_H$ have been respectively
fixed to the same optical redshift references as used in the original XBACS
database, and to the weighted means of the $N_H$ measurements from the
\cite{hilx:DL90} database included within $1 \deg$ of the cluster center.
The resulting temperatures $kT$, abundance relative to solar composition $\mu$
and normalization $S_0$ are then reported. These measurements come from
spectra extracted inside a circle of $400~arcsec$ radius, and hence integrate
any central excess emission; that is why temperatures may appear a bit low.

\subsection*{Direct temperature profiles}
Spectra have also been extracted within $40~arcsec$-wide concentric annuli
following the same algorithm than for global spectra. Fitting them using
the same XSPEC settings hence led to a direct measure of the temperature
profile of each cluster. However, no PSF correction have been applied from
one annulus to another thus probably leading to a general underestimation
in the fitted temperatures, specially toward the center (\cite{hilx:Mark02}).

A last spectrum has been extracted and fitted, in the same region than that
used for the \bm\ fitting, {\it i.e.} an annulus starting at $40~arcsec$ and
extending upto the arbitrary limit of $400~arcsec$. This allows to compare
between the global temperature with the one found outside the core region,
which may host a cooling flow.

\subsection*{Intrinsic parameters calculation}
The luminosity distance $D_L$ and $1~arcsec$ equivalent length $R_{1"}$ at
the cluster position are deduced from the cluster redshift $z$ and the
default cosmology ($\Omega_0=1$, $\Omega_\Lambda=0$ and
$H_0=50~h_{50}~km/s/Mpc$).

The virial radius $R_V$ has been identified with the distance $r_{200}$ from
the cluster center where the mean enclosed overdensity equals $200$, as
calibrated upon the measured X-ray temperature by \cite{hilx:EMN96}.

The X-ray luminosity $L_X$ was integrated between our \xmm\
energy bandwith ($0.3 - 6.3~keV$) within the virial radius defined above. We
applied a numerical bolometric correction to deduce the absolute luminosity
$L_X(bol)$.

The galaxies velocities dispersions $\sigma_{wl_D}$ and $\sigma_{wl_B}$ have
been reprinted from two weak lensing analysis, respectively from
{\it NOT/ALFOSC} and {\it UH8k} detectors (\cite{hilx:Dahle02}), and
{\it CFHT12k} instrument (\cite{hilx:Bardeau02}). The corresponding $\beta$
parameters were computed according to the relationship with the measured X-ray
temperature in the isothermal hydrostatic equilibrium case (\cite{hilx:CFF76}).

The electronic density $n_e(0)$ at the center of the cluster has been estimated
using the standard isothermal \bm\ equations (\cite{hilx:CFF76}) and the
previously measured X-ray temperature and luminosity. By integration within
the virial radius defined above, we also estimated the cluster gas mass.
Finally, the total virial mass has been deduced from two different calibrated
relations upon the measured X-ray temperature ($M_B$ and $M_T$ resp. from
\cite{hilx:CFF76,hilx:AE99}) as well as from the virial theorem relation upon
the galaxies velocities dispersion ($M_V$ from the $\sigma_{wl_B}$ defined
above, except for A773 for which we used $\sigma_{wl_D}$). The gas fraction
$f_g$ simply is the ratio of the gas mass to the total mass (we chose $M_B$
as the isothermal reference).

\section*{DISCUSSION}

\begin{figure}
  \centering
  \includegraphics[width=75mm]{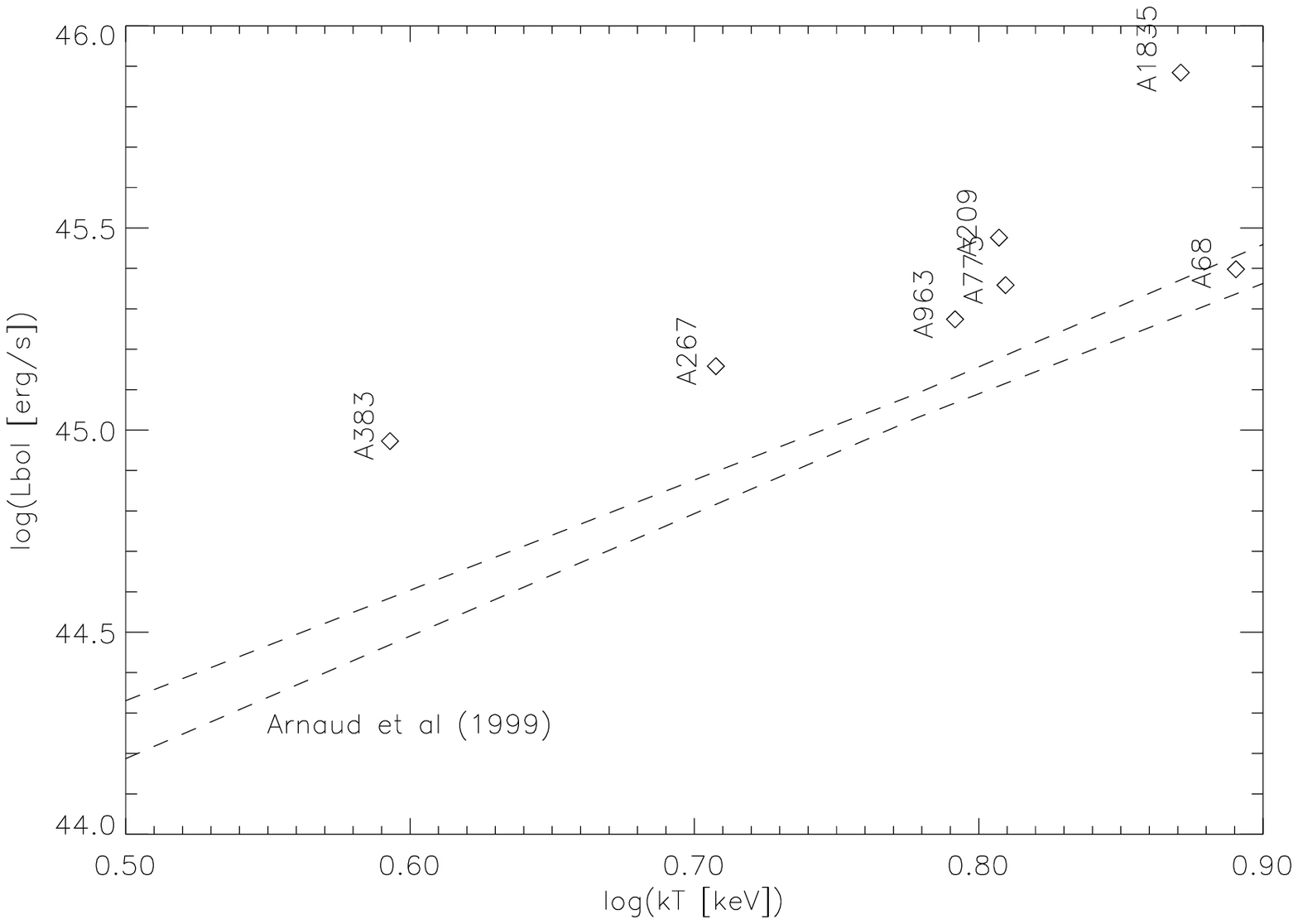}
  \includegraphics[width=75mm]{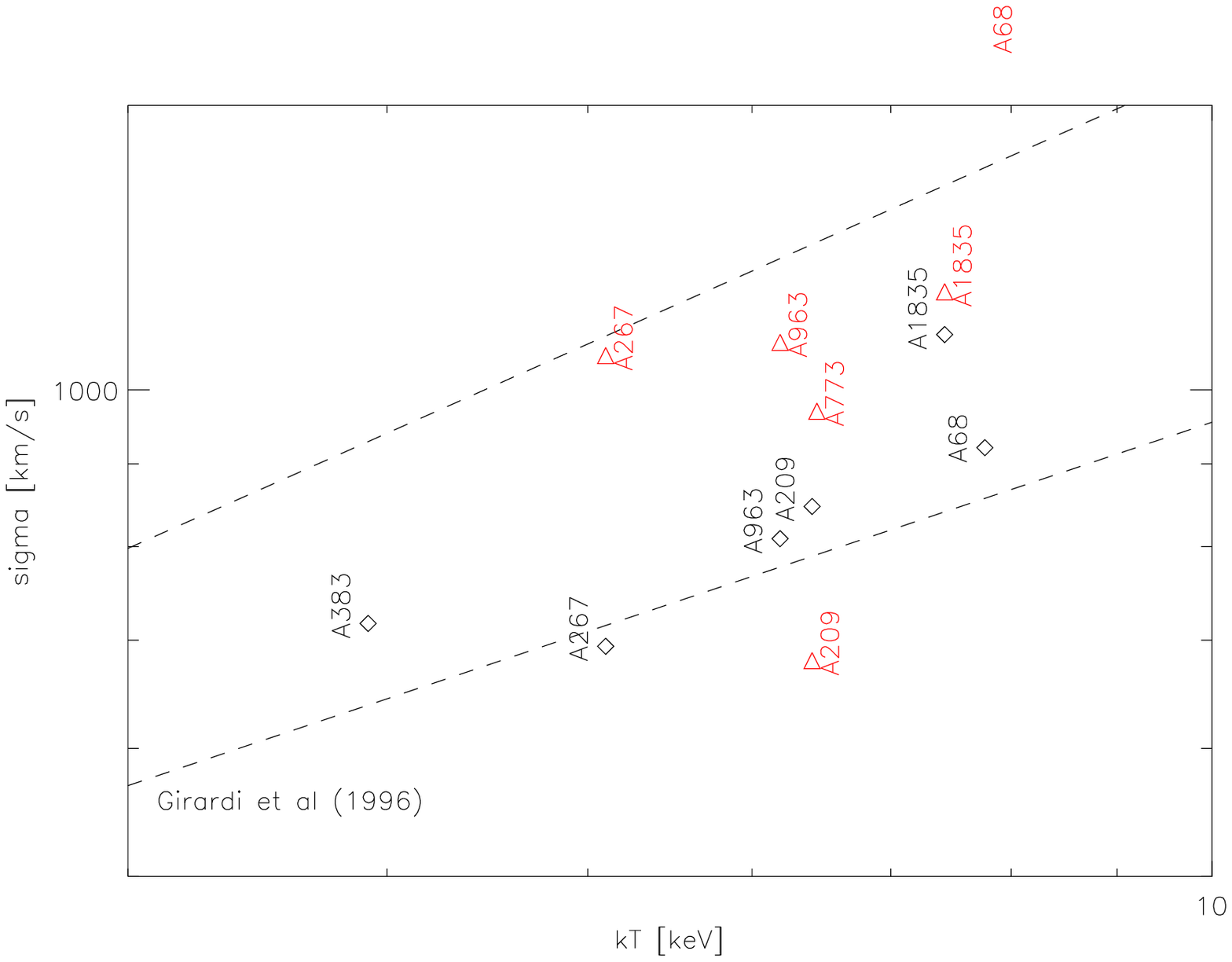}
  \caption{\label{hilx:lumi}Left: temperature-luminosity diagram as compared
to the $90\%$ confidence level range from \cite{hilx:AE99} model.
Right: temperature-velocity dispersion diagram as compared
to the $90\%$ confidence level range from \cite{hilx:Girard96} model
(diamonds show values from \cite{hilx:Bardeau02}, triangles from
\cite{hilx:Dahle02}).}
  \includegraphics[width=75mm]{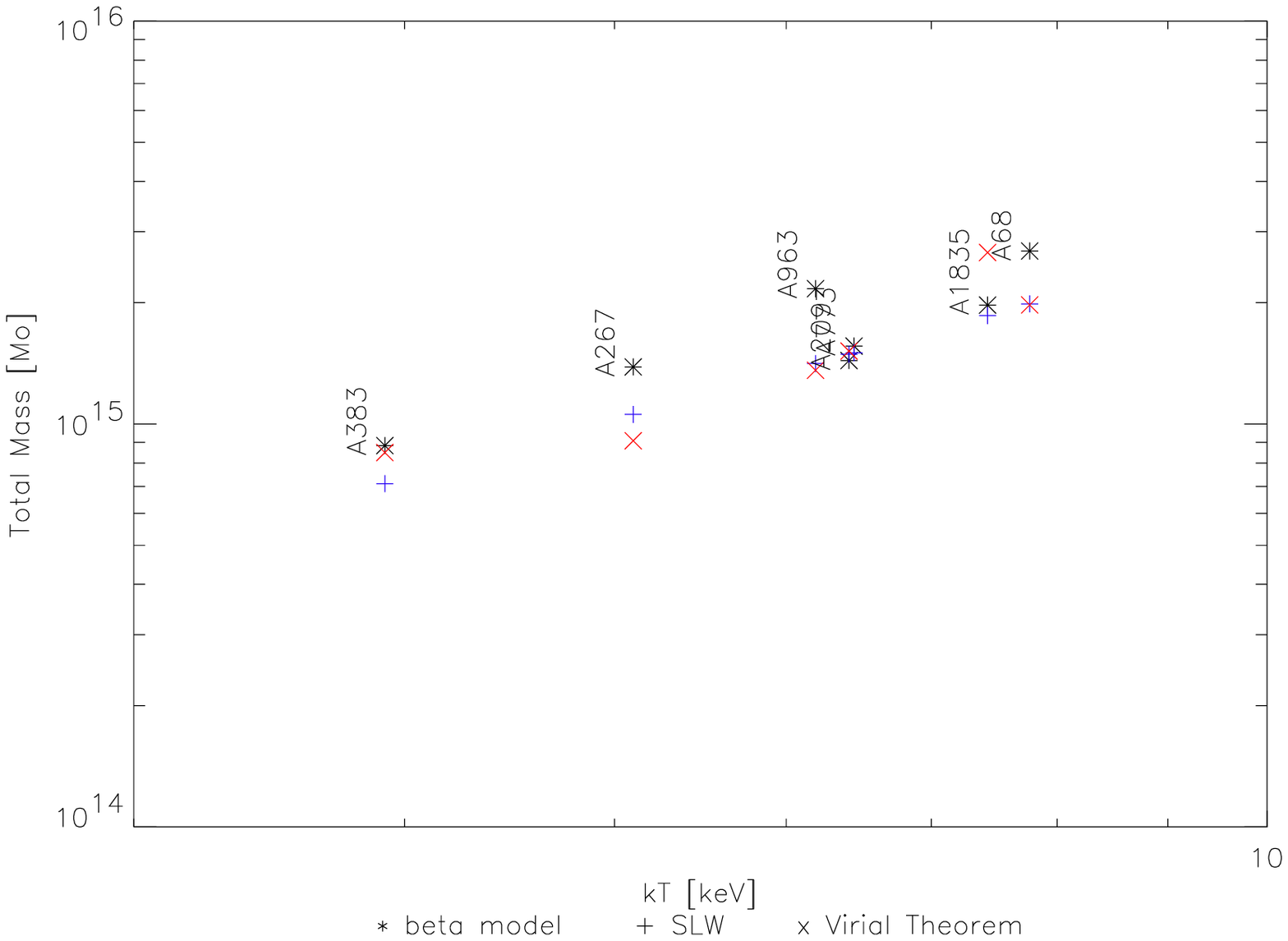}
  \includegraphics[width=75mm]{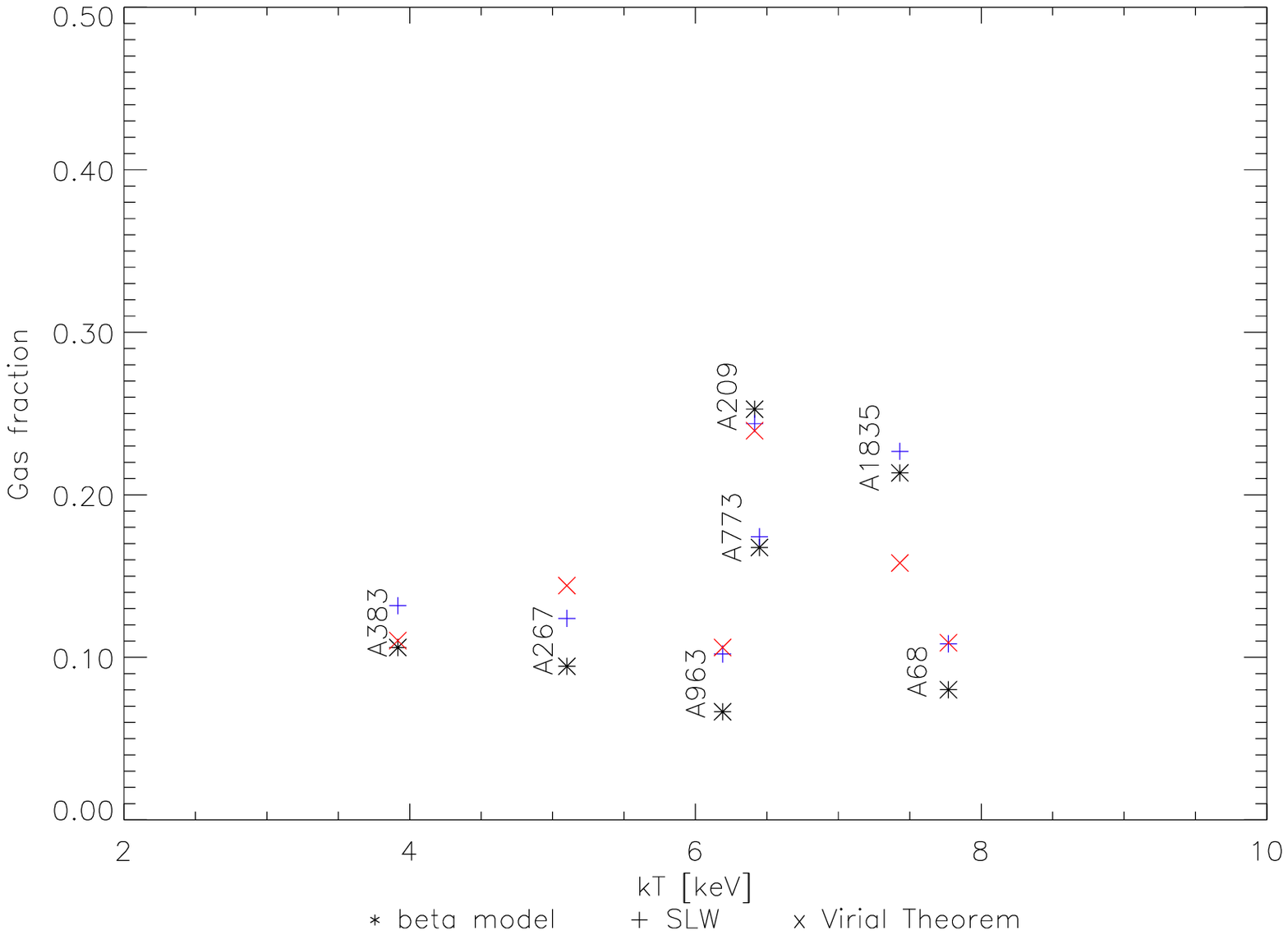}
  \caption{\label{hilx:mass}Left: temperature-mass diagram, for different
mass estimators, showing a mean power index of $1.576$. Right: gas mass
fraction, for the same estimators, as a function of the X-ray temperature.}
  \includegraphics[width=75mm]{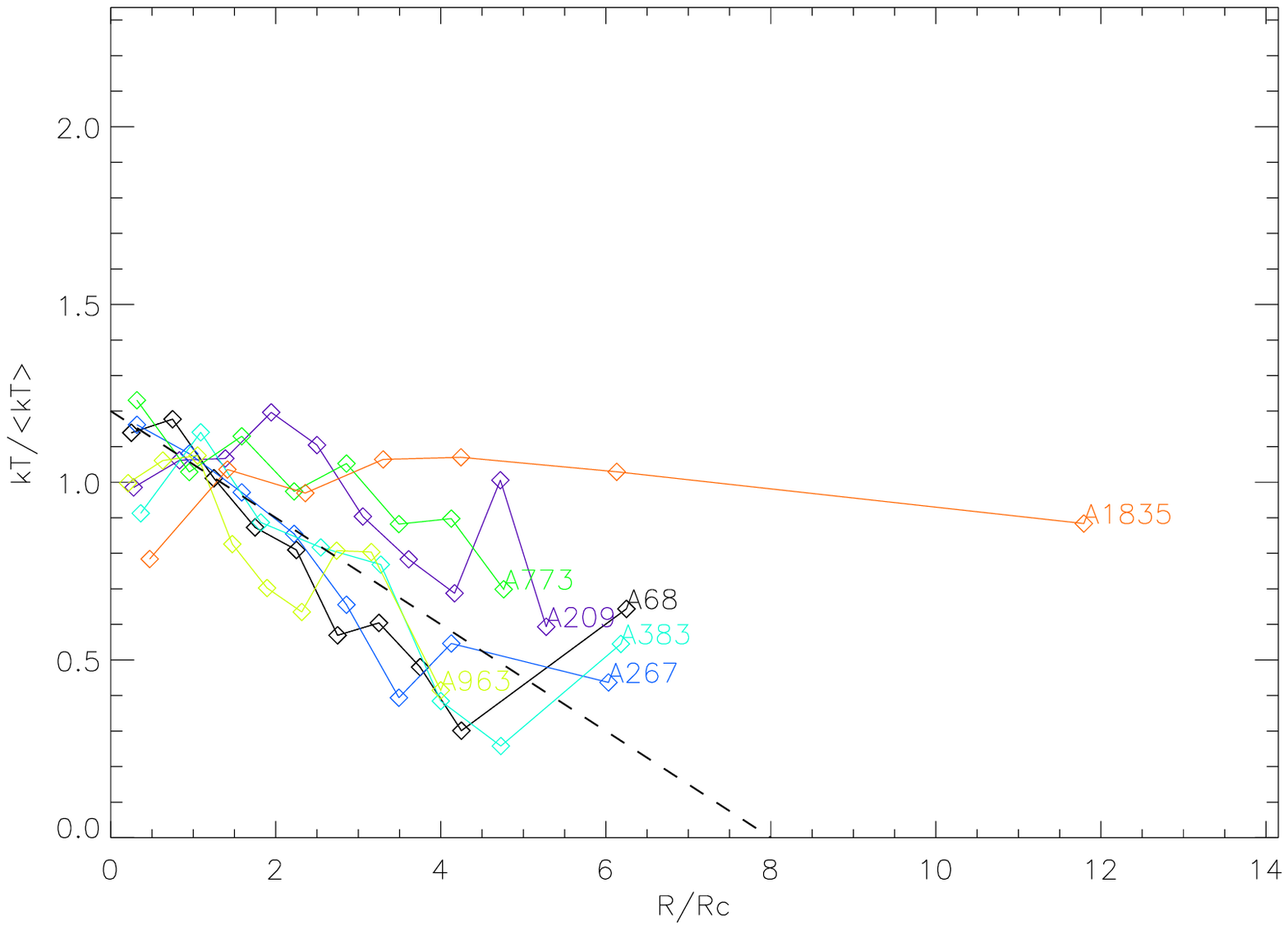}
  \includegraphics[width=75mm]{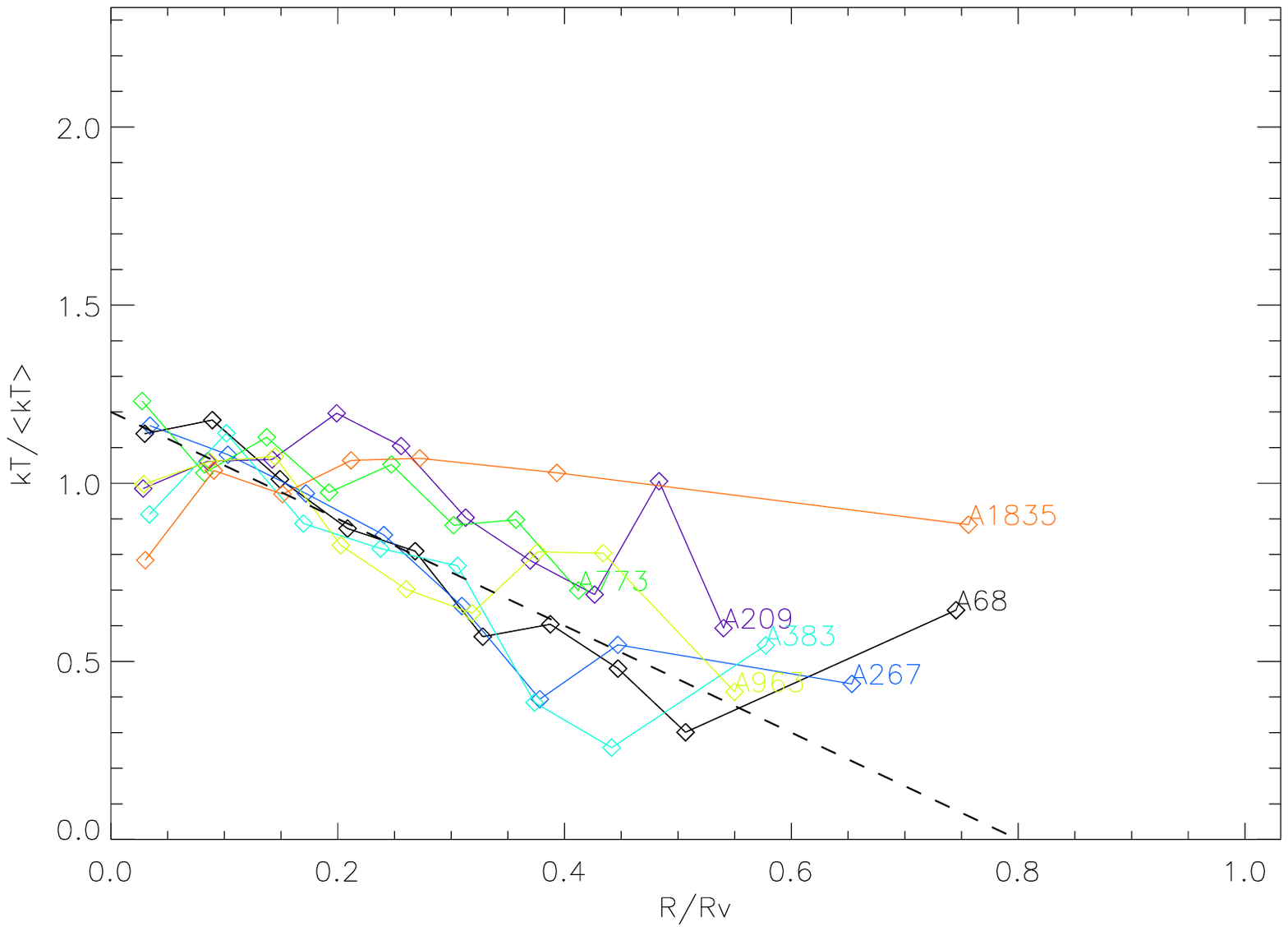}
  \caption{\label{hilx:prftmp}Temperature profiles with radius rescaled
either to core radius or virial radius units. The dashed line stands for the
average slope in the central part of the polytropic simulation of
\cite{hilx:Loken02}.}
\end{figure}

All analysed clusters show an excess in central brightness that seems
correlated with the presence of a very bright central galaxy in optical images.
They hence are all candidates for cooling flow hosts, but A267 and A383
temperature profiles do not exhibit significant cooling cores. In addition,
all clusters but A1835 tend to follow the universal temperature profile
calibrated by \cite{hilx:Loken02} (\fig{hilx:prftmp}), radii being rescaled
either to core radius or virial radius natural units, with a declining slope
toward the edges. They are also very close to the \cite{hilx:DeGrandi02}
polytropic interpretation of their {\it Beppo-SAX} observed clusters data,
which show slopes more horizontal at high radii for those hosting a cooling
flow than for non-cooling flow specimens, although our sample does not present
a steeply cooling core in average.

If the problem of the cooling core may be bypassed by using only outer regions
for temperature and \bm\ fitting, these declining profile still questions the
hypothesis of isothermality followed throughout this first analysis.
Nevertheless, we expect from the general good agreement of the different mass
(\fig{hilx:mass})
and $\beta$ parameters estimations that our optical ({\it CFHT12k}) and X-rays
(\xmm) measurements are still both quite consistent with the isothermal model.
But we note also that all calculations rely on a virial radius that has been
estimated using a relation based on the isothermal hypothesis; that illustrates
the need for another independant estimator of this radius. Even the clue
given by the temperature profiles (\fig{hilx:prftmp}), that seem to indicate
that the virial radius is about $10$ times the core radius, is not so helpful
if models fitted to either optical or X-ray data still rely upon isothermality.

We plotted a temperature-luminosity logarithmic diagram (\fig{hilx:lumi}) for
our subsample and compared it to the model proposed by \cite{hilx:AE99}. We
observe that five clusters are indeed well co-aligned but following a shallower
slope. A209 seems to depart from this alignment, as well as A1835 in a very
significant manner.
We also compared a temperature-velocity dispersion diagram with the model
from \cite{hilx:Girard96} and found it to be very close to the lower $90\%$
confidence level boundary, with again the exception of A1835.

Finally, the temperature-mass diagram is remarkably consistent with a mean
power index of $1.576$, while the gas mass fraction seems to average about
$10\%$ except again for A1835, as well as A209 and A773.

\section*{CONCLUSIONS \& PERSPECTIVES}

While early \xmm\ observations (\cite{hilx:Arnaud01b,hilx:Maj02}) did not
confirm declining temperature profiles as seen by late {\it ASCA} observations
(\cite{hilx:Mark98}) and {\it Beppo-SAX} (\cite{hilx:DeGrandi02,hilx:Loken02}),
we find in the present analysis of 7 XBACS clusters that 6 of them do follow
this kind of polytropic model. The seventh, A1835, systematically departs from
the average in all diagrams (temperature-mass, temperature-luminosity), which
confirms the rather complicated nature of its internal dynamics
(\cite{hilx:Maj02}).
 
However, we still are working toward the comparison of these emission weighted
temperature profiles, that have been obtained without correction for PSF
and may thus have been underestimated (\cite{hilx:Mark02}), with profiles
built on the basis of hardness ratio maps, that may be adaptively handled to
account for the PSF and binned between concentric isophotes rather than
concentric annulii to adapt to non-spherical morphologies. Also, more work is
intended so as to derive accurate
error bars on each diagram, and attempt to estimate the virial radii from
independent optical datasets.

Finally, this work has been made possible through the use of a pipeline
analysis environment, specifically aimed at dealing with extended sources
data from \xmm\ instruments. It has confirmed the importance of maintenance
and upgrades, to keep up-to-date with the instruments calibration database, as
well as the need for such automatic tools in the purpose of building catalogs
from the latest X-ray telescopes, so that massive cluster investigation may be
possible in the future: access to large statistical samples, cross-compare
mass and morphological measurements with optical data, correlate X-ray and
Sunyaev-Zel'Dovich data\ldots The ESA's {\it Planck} satellite is promising
thousands of millimetric clusters detections by the end of the decade, which
could lead to narrower constraints on cosmology, provided an extensive X-rays
database, like BAX (\cite{hilx:Sadat02}), also exists.

\section*{ACKNOWLEDGEMENTS}

This work is based on observations obtained with \xmm, an {\em ESA} science
mission with instruments and contributions directly funded by {\em ESA} Member
States and the {\em NASA}. This works also benefits from observations conducted
with the CFH12k camera at the Canada France Hawaii Telescope.

The authors wish to thank Bruno Altieri, Christian Erd, David Lumb, Richard
Saxton, as well as the whole ``epic-cal'' team for their help. Our thanks also
go to Pasquale Mazzotta for fruitful discussions and comparisons between
results from {\it Chandra} and \xmm\ observatories.

J.P.K. acknowledges support from an ATIP/CNRS grant and from PNC.
I.R.S. acknowledges support from the Royal Society and Leverhulme Trust.

E-mail address of P.B. Marty: \url{philippe.marty@ias.u-psud.fr}


\end{document}